# Effects of interaction between nanopore and polymer on translocation time

Mohammadreza Niknam Hamidabad and Rouhollah Haji Abdolvahab
Physics Department, Iran University of Science and Technology (IUST), 16846-13114, Tehran, Iran
rabdolvahab@gmail.com
+982173225887

**Abstract**: Here using LAMMPS molecular dynamics (MD) software, we simulate polymer translocation in 2 dimensions. We do the simulations for weak and moderate forces and for different pore diameters. Our results show that in both non-equilibrium and equilibrium initial conditions, translocation time will always increase by increasing binding energy and or increasing pore diameter. Moreover, scaling exponent of time versus force is -0.9531 in accordance to our predecessors. The comparison between equilibrium and non-equilibrium initial condition shows that the translocation time is very sensitive to the initial condition. Translocation time of the relaxed polymers for interaction energy of $8k_BT$ is smaller from the non-equilibrium case even in the small energy of $1k_BT$.

## 1-Introduction

Biopolymer's translocation through nanopores is a ubiquitous process in both biology and biotechnology. Protein translocation through membranes, RNA translocation through nucleopores and RNA injection through host cell by a virus are some instances [1, 2]. Moreover, it has technological applications such as drug delivery [3] and DNA sequencing [4]. Consequently, polymer translocation through the nanopore is one of the most active fields in biophysics and soft matter [5, 6].

*In vivo* Chaperone-assisted polymer translocation is one of the important deriving mechanism. In this method, which is used in cells [7], as soon as the polymer went through the Trans side, a protein, called chaperone, bound to it and prevent the polymer from backsliding [7-12]. However, *in vitro*, among the methods of polymer translocation through a nanopore, translocation deriving by external force is one of the most common experimental and computational approaches [6, 13-19].

In translocation deriving by an external force, different parameters like, external force, nanopore length, polymer length, crowding *etc*. on translocation time are investigated. One of the important parameters here is binding energy between channel wall and polymer, which we consider in this work.

In our simulation, we investigate polymer translocation in two different initial conditions, namely, equilibrium and non-equilibrium in at least 8 different interaction energies. Moreover, we change the pore diameter and consider 3 different amount of it. Simulation results indicate an increase in translocation time by increasing interaction energy and/or pore diameter in both cases. In what follows, we first outline our theoretical model. Then we report and analyze the results from our simulations. We finished our work by a conclusion based on the simulation results.

## 2-Theoretical model

We model the polymer using a chain of masses and springs so that there is a finitely extensible nonlinear elastic (FENE) potential (Eq. 1) between adjacent monomers. There is also a Leonard-Jones potential (Eq. 2) between all monomers.

$$U_{FENE} = -\frac{1}{2} k R_0^2 \ln\left(1 - \frac{r^2}{R_0^2}\right) \tag{1}$$

$$U_{LJ} = \begin{cases} 4\varepsilon\left[\left(\frac{\sigma}{r}\right)^{12} - \left(\frac{\sigma}{r}\right)^6\right] + \varepsilon & r \leq r_{cut} \\ 0 & r > r_{cut} \end{cases} \tag{2}$$

$R_0$ is the maximum allowed distance between adjacent monomers and $k$ is the spring constant. $\sigma$, shows the monomer size, $\varepsilon$ is the potential depth and $r_{cut}$ is the effective radius of the Leonard-Jones potential. The Leonard-Jones potential is also used for pore walls with a different $r_{cut}$.

We use of Langevin dynamics for the simulation. In this method for each monomer, one can write:

$$m\ddot{\vec{r}}_i = \vec{F}_i^C + \vec{F}_i^F + \vec{F}_i^R \tag{3}$$

in which $m$ is the monomer mass, $F_i^C$ calls for conservative forces, $F_i^F$ is the friction and $F_i^R$ shows stochastic force on the monomer. $F_i^R$ is related to monomer's velocity as:

$$\vec{F}_i^F = -\xi \vec{V}_i \tag{4}$$

where $\xi$ is the friction constant.
For conservative force:

$$\vec{F}_i^C = -\vec{\nabla}(U_{LJ} + U_{FENE}) + \vec{F}_{external} \tag{5}$$

$\vec{F}_{external}$ is the external force on the polymer through nanopore which we define:

$$\vec{F}_{external} = F\hat{x} \tag{6}$$

where the direction of the force coincides with the axis of the nanopore through the Trans side.

## 2-1-Initial configuration and simulation parameters

Nanopore has a length of $6\sigma$, where $\sigma$ is the size of a monomer and we consider pores of diameters 3, 4 and $5\sigma$. We put the first monomer at the end of the nanopore. We situate the other monomers of the polymer in the equilibrium state with respect to each other but in a straight line in front of the pore. In non-equilibrium initial condition, we just start the simulation while in equilibrium initial condition the polymer have enough time to equilibrate and then the translocation starts. In equilibration of the second initial state, we fix the monomers through the channel. This part of simthe ulation, takes of around 20% (for the slowest polymers) to 40% (for the fastest polymer) of the whole simulation time. In order to decrease the error of the mean translocation time, we repeat the process for at least 1000 times.

We use from gyration radius to find the equilibration. We see the gyration radius change over time and continue the simulation until it reaches to its saturation value with small fluctuations.

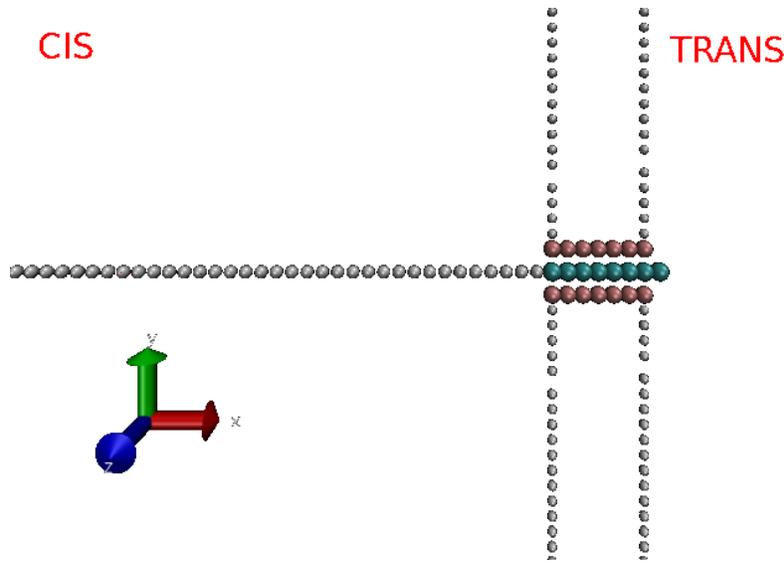

Figure 1: Initial configuration for a non-equilibrium sample with a pore diameter of $5\sigma$ and $N = 50$.

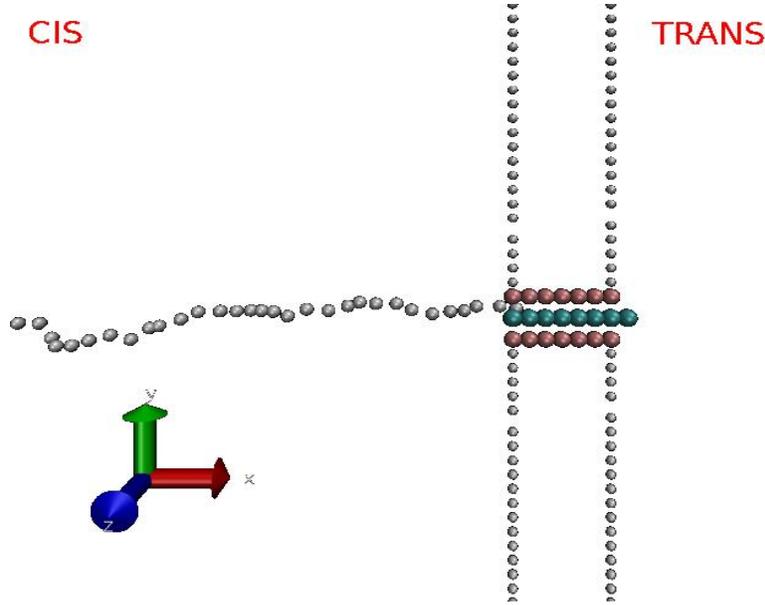

Figure 2: Initial configuration for an equilibrium sample with a pore diameter of $5\sigma$ and $N = 50$.

Time unit of the simulation is obtained from:
$$t_{LJ} = \left(\frac{m\sigma^2}{\varepsilon}\right)^{\frac{1}{2}} \tag{7}$$

This time unit is of the order of $10 fs$.
Here we use two different external forces of weak and moderate for simulation. The moderate force satisfies the following relation [15]:

$$\frac{k_B T}{\sigma N^\nu} \leq F \leq \frac{k_B T}{\sigma} \tag{8}$$

in which, $\nu$ is the Flory exponent and $N$ is the total number of monomers. External forces that we used as weak and moderate forces are 3.5pN and 6.5pN, respectively. Effective interaction radius for inthe teraction of polymer and nanopore take as $2\sigma$ and for other interactions $2^{1/6}\sigma$. By setting the cut off distance $r_{cut} = 2^{1/6}\sigma$, to exclude attractive interactions, we models the polymer to be immersed in a good solvent [10]. Moreover, we take the parameter of energy (apart from polymer and nanopore) as $\varepsilon = k_B T$. The length is of order angstrom. Friction coefficient equals $\xi = 0.7 \frac{m}{t_{LJ}}$. Parameters of FENE potential taken to be $k = 40 \frac{\varepsilon}{\sigma^2}$, $R_0 = 1.5\sigma$ and the mass of each monomer $m = 70 amu$ [20].

# 3-Results and discussion

Due to two states of different initial conditions of equilibrium and non-equilibrium, we present our results accordingly.

## 3-1-Non-equilibrium state

Increasing binding energy will always increase the translocation time in non-equilibrium initial condition. Moreover, our simulation results reveal that increasing pore diameter from 3 to 4 and then $5\sigma$ will always increase the mean translocation time of the polymer in both weak and moderate forces (figures 3, 4).

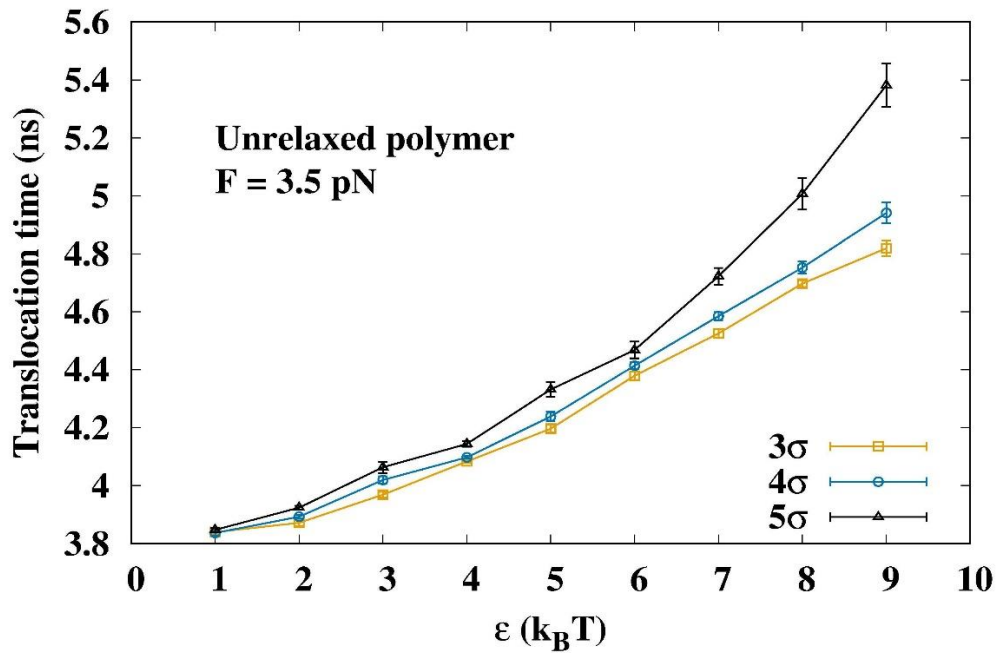

Figure 4: Mean translocation time of the polymer in the non-equilibrium state against interaction energy in 3 different pore size of 3, 4 and $5\sigma$ and for the weak force of $3.5 pN$. The mean is taken over 1000 sample.

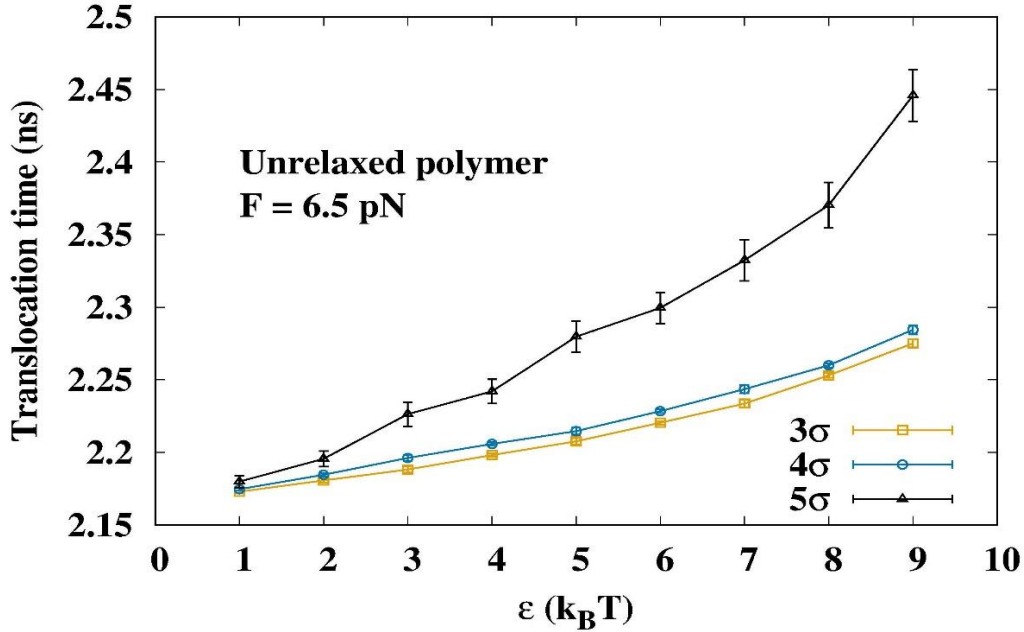

Figure 3: Mean translocation time of the polymer in the non-equilibrium state against interaction energy in 3 different pore size of 3, 4 and $5\sigma$ and for the moderate force of $6.5 pN$. The mean is taken over 1000 sample.

Comparison of figure 3 and figure 4 shows that the mean translocation time of the polymer in weak force is significantly larger than translocation time in the moderate force. This increase is expected due to decreasing the driving force. Moreover, the difference between translocation time of the polymer in case of pore diameter $5\sigma$ and to smaller pores increase by increasing driving force from weak to moderate.
Polymer translocation time will decrease by increasing the external force. This change in moderate forces helps us to calculate a scaling exponent. We plot the translocation time versus force for moderate forces in the log-log axis in figure 5. We measure the scaling exponent of translocation time versus forces using the process of curve fitting as -0.9757. In these simulations the interaction energy $\varepsilon = 6$, pore diameter was $4\sigma$ and polymer length $N = 50$. It has been calculated in literature only for equilibrium initial state [14, 16-18].

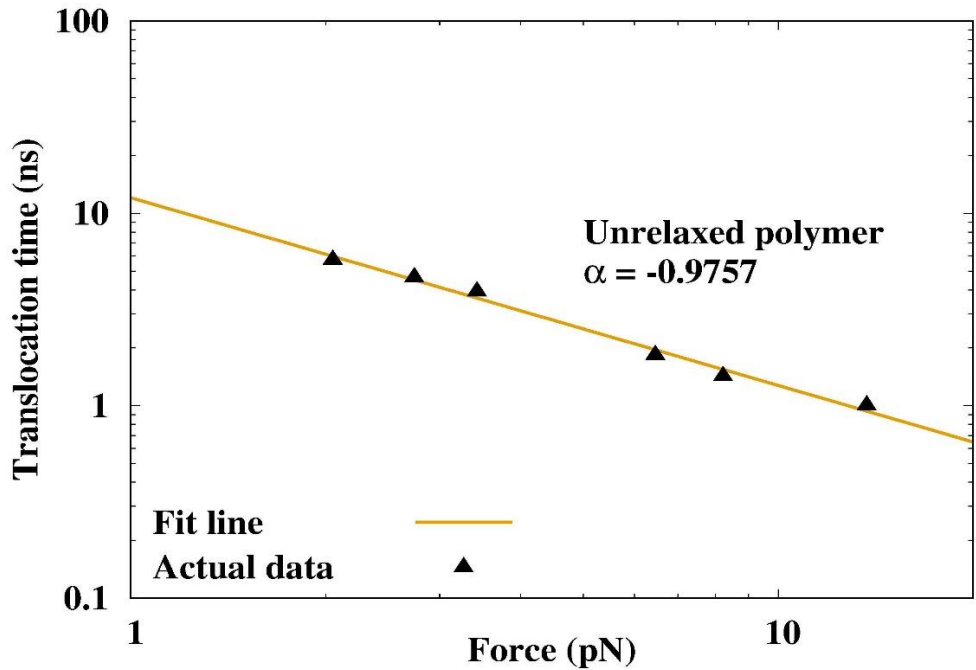

Figure 5: Polymer translocation time against moderate forces in a log-log plot. The measured scaling exponent is -0.9757.

### 3-2-equilibrium state

In equilibrium initial configuration, the results for weak forces are similar to that of non-equilibrium state. Increasing both the binding energy and pore diameter, will increase the translocation time and slow down the translocation process (figure 6). An interesting result here is the slope of the curves. The slope of the curve for the diameter of size $5\sigma$ is significantly greater than that of pore diameters of 3 and $4\sigma$.

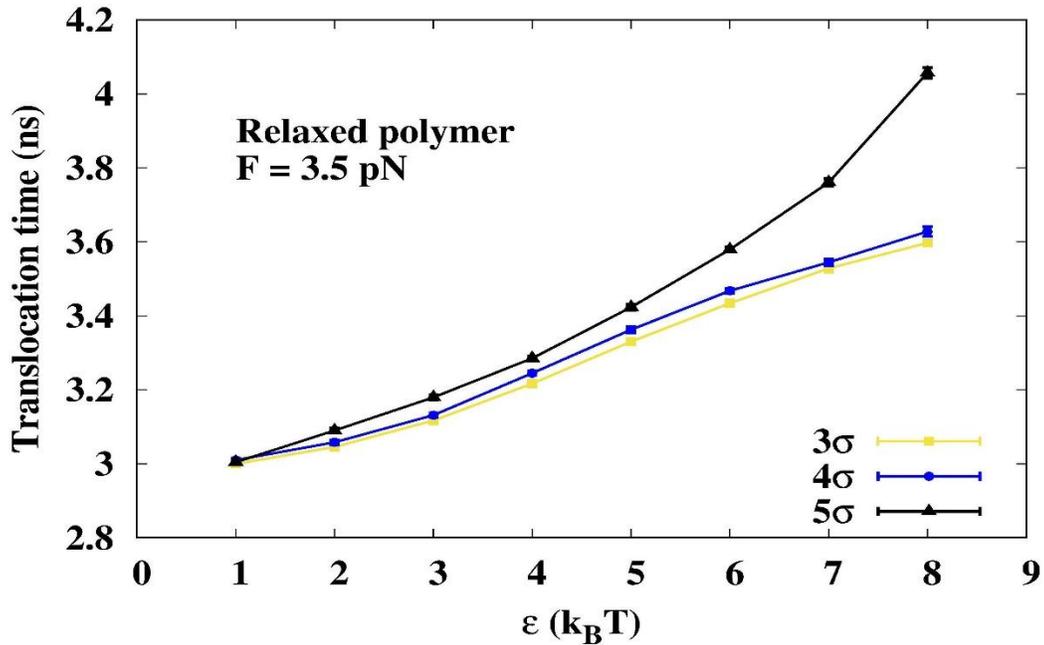

Figure 6: Mean translocation time of the polymer in equilibrium state against interaction energy in 3 different pore size of 3, 4 and 5$\sigma$ and for the weak force of 3.5$pN$. The mean is taken over 1000 sample.

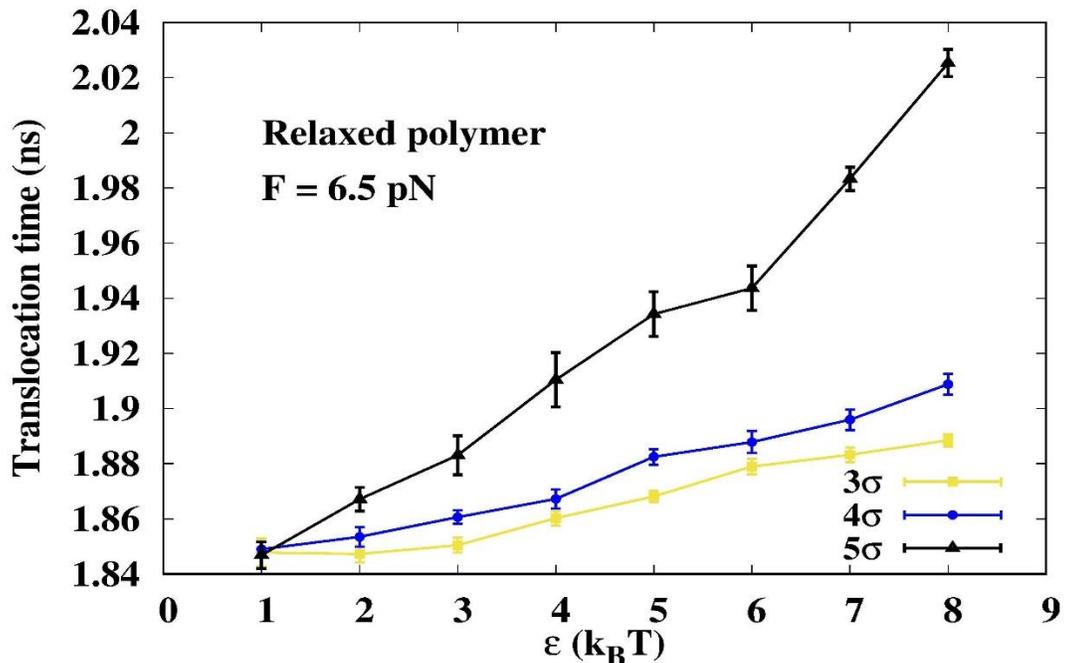

Figure 7: Mean translocation time of the polymer in equilibrium state against interaction energy in 3 different pore size of 3, 4 and 5$\sigma$ and for the weak force of 6.5$pN$. The mean is taken over 1000 sample.

We change the external force in the simulation and calculate the translocation time. Log-log plot of translocation time against forces is sketched in figure 8. Curve fitting on the figure shows that the scaling exponent of translocation time versus forces is equal to -1.008. Here the interaction energy is taken to be $\varepsilon = 6$, pore diameter was $4\sigma$ and polymer length $N = 50$. Luo *et al.* in 2009 report the exponent 1 for weak and moderate forces[18]. Huopaniemi *et al.* in 2007 calculate the exponent for weak, moderate and strong forces as -0.94[14], which are in good agreement with our calculations.

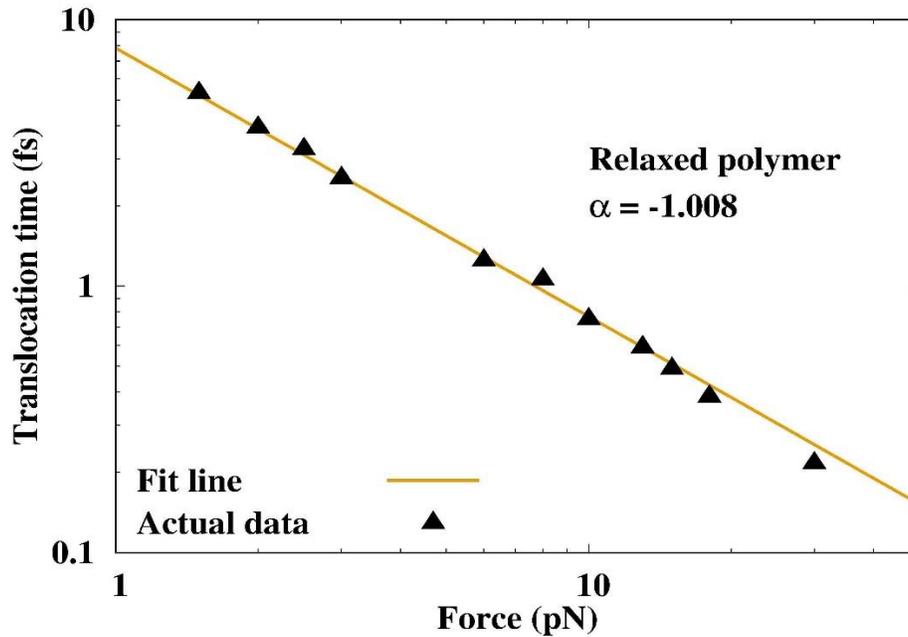

Figure 8: Polymer translocation time against force in a log-log plot. The measured scaling exponent is -1.008.

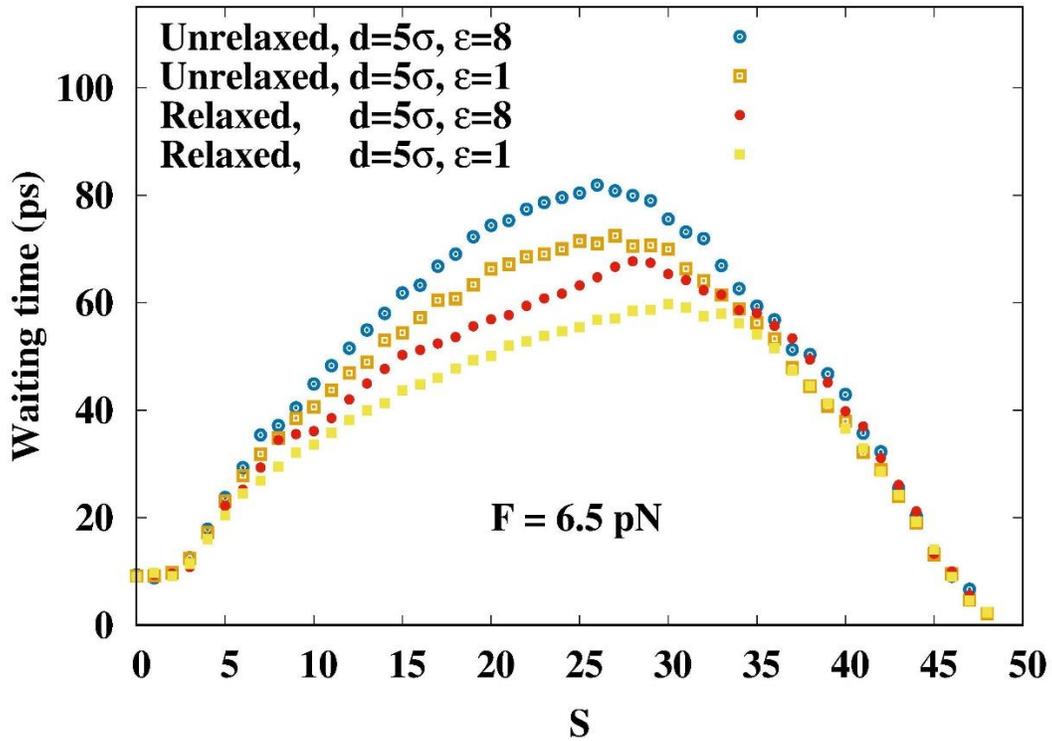

Figure 9: Plots of waiting times for both initial conditions and in 2 different binding energies 1 and 8, for pore of diameter size $5\sigma$.

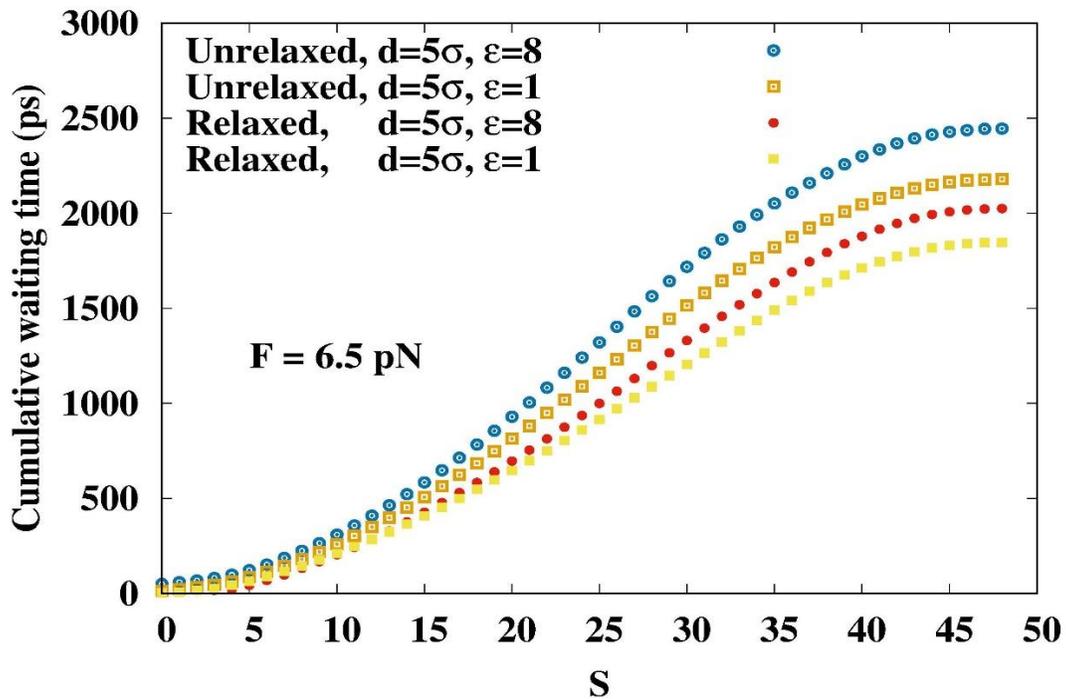

Figure 10: Plots of cumulative waiting times for both initial conditions and in 2 different binding energies 1 and 8, for pore of diameter size $5\sigma$.

Waiting time of the polymer translocation for pore diameter $5\sigma$, in both initial conditions and for to binding energies of 1 and 8 is plotted in figure 9. It shows 3 stages of polymer translocation. Due to the empty space in the right the translocation of the first monomers are fast. Afterward, the translocation becomes slower and slower until, nearly, the middle of the polymer. Then the translocation's speed increase until the end. Cumulative of the above waiting time is shown in the figure 10. As expected from the waiting times, most of the separation of translocation times take place in the middle of the translocation.

## Conclusion

We studied polymer translocation through a nanopore using a series of MD simulations. An external force derives the translocation. We investigated especially the effects of interaction between polymer and pore wall in our coarse-grained simulation.

Our simulation results show that in both initial conditions, the translocation time will be increased by increasing the interaction energy and/or pore diameter. However, the initial condition is very important. As an example the translocation time for the unrelaxed polymer in binding energy of 1 is larger than the translocation time for the relaxed polymer with energy 8.

The results also show that by increasing the pore diameter from $3\sigma$ to $5\sigma$ the difference between translocation times will be increased. This result may be used to separate different polymers using their translocation time.

**Acknowledgement**: We use from LAMMPS molecular dynamic software for the simulation [21]. We also used from VMD for graphic representation [22] and Gnu plot for plotting the figures [23].